\documentclass[conference]{IEEEtran}
\IEEEoverridecommandlockouts

\usepackage{cite}
\usepackage{amsmath,amssymb,amsfonts}
\usepackage{algorithmic}
\usepackage{graphicx}
\usepackage{textcomp}
\usepackage{xcolor}
\usepackage{array}
\usepackage{url}
\usepackage{comment}
\usepackage[symbol]{footmisc}
\def\BibTeX{{\rm B\kern-.05em{\sc i\kern-.025em b}\kern-.08em
    T\kern-.1667em\lower.7ex\hbox{E}\kern-.125emX}}
    
\newcommand{\indeg}{d^{\mathrm{in}}}
\newcommand{\outdeg}{d^{\mathrm{out}}}

\begin{document}

\title{A Highly Parallel FPGA Implementation\\ of Sparse Neural Network Training\thanks{An abridged version of this work was accepted as a short paper at ReConFig: 2018 International Conference on Reconfigurable Computing and FPGAs. This is the full version of this work.}}

\author{\IEEEauthorblockN{Sourya Dey, Diandian Chen, Zongyang Li, Souvik Kundu, \\Kuan-Wen Huang, Keith M.~Chugg and Peter A.~Beerel}
\IEEEauthorblockA{\textit{Ming Hsieh Department of Electrical Engineering} \\
\textit{University of Southern California}\\
Los Angeles, California 90089, USA \\
\{souryade, diandiac, zongyang, souvikku, kuanwenh, chugg, pabeerel\}@usc.edu}
}

\maketitle

\begin{abstract}
We demonstrate an FPGA implementation of a parallel and reconfigurable architecture for sparse neural networks, capable of on-chip training and inference. The network connectivity uses pre-determined, structured sparsity to significantly reduce complexity by lowering memory and computational requirements. The architecture uses a notion of edge-processing, leading to efficient pipelining and parallelization. Moreover, the device can be reconfigured to trade off resource utilization with training time to fit networks and datasets of varying sizes. The combined effects of complexity reduction and easy reconfigurability enable significantly greater exploration of network hyperparameters and structures on-chip. As proof of concept, we show implementation results on an Artix-7 FPGA.
\end{abstract}


\section{Introduction}\label{intro}

Neural networks (NNs) in machine learning systems are critical drivers of new technologies such as image processing and speech recognition. Modern NNs are built as graphs with millions of trainable parameters \cite{Krizhevsky2012,Szegedy2015,He2016}, which are tuned until the network converges. This parameter explosion demands large amounts of memory for storage and logic blocks for operation, which make the process of training difficult to perform \emph{on-chip}. As a result, most hardware architectures for NNs perform training off-chip on power-hungry CPUs/GPUs or the cloud, and only support inference capabilities on the final FPGA or ASIC device \cite{Chen2014DN,Chen2015,Han2016EIE,Zhou2016,Yufei2017,Han2017ESE,Wang2018}. 
Unfortunately, off-chip training results in a non-reconfigurable network being implemented on-chip which cannot support training time optimizations over model architecture and hyperparameters. This severely hinders the development of \emph{independent NN devices} which a) dynamically adapt themselves to new models and data, and b) do not outsource their training to costly cloud computation resources or data centers which exacerbate problems of large energy consumption \cite{Shehabi2016}.

Training a network with too many parameters makes it likely to overfit \cite{Denil2013}, and memorize undesirable noise patterns \cite{Zhang2016_2}. Recent works \cite{Dey2017_ICANN,Dey2018_ITA,Aghasi2017,Ullrich2017} have shown that the number of parameters in NNs can be significantly reduced without degradation in performance. This motivates our present work, which is to train NNs with reduced complexity and easy reconfigurability on FPGAs. This is achieved by using \emph{pre-defined sparsity} \cite{Dey2017_ICANN,Dey2017_Asilomar,Dey2018_ITA}. Compared to other methods of parameter reduction such as \cite{Chen2015,Srivastava2014,Han2016DC,Gong2014,Wang2018}, pre-defined sparsity does not require additional computations or processing to decide which parameters to remove. Instead, most of the weights are always absent, i.e. sparsity is enforced \emph{prior to training}. This results in a sparse network of lesser complexity as compared to a conventional fully connected (FC) network. Therefore the memory and computational burdens posed on hardware resources are reduced, which enables us to accomplish training on-chip. 
Section \ref{arch} describes pre-defined sparsity in more detail, along with a hardware architecture introduced in \cite{Dey2017_ICANN} which exploits it. 

A key factor in NN hardware implementation is finite bit width effect. A previous FPGA implementation \cite{KaanK2017} used fixed point adders, but more resource-intensive floating point multipliers and floating-to-fixed-point converters. Another previous implementation \cite{Suyog2017} used probabilistic fixed point rounding techniques, which incurred additional DSP resources. Keeping hardware simplicity in mind, our implementation uses only fixed point arithmetic with clipping of large values.

The major contributions of the present work are summarized here and described in detail in Section \ref{fpga}:
\begin{itemize}
    \item The first implementation of NNs which can perform both training and inference on FPGAs by exploiting parallel edge processing. The design is parametrized and can be easily reconfigured to fit on FPGAs of varying capacity.
    \item A low complexity design which uses pre-defined sparsity while maintaining good network performance. To the best of our knowledge, this is the first NN implementation on FPGA exploiting pre-defined sparsity.
    \item Theoretical analysis and simulation results which show that sparsity leads to reduced dynamic range and is more tolerant to finite bit width effects in hardware.
\end{itemize}

\section{Sparse Hardware Architecture}\label{arch}

\subsection{Pre-defined Sparsity}\label{pds}
Our notation treats the input of a NN as layer 0 and the output as layer $L$. The number of neurons in the layers are $\{N_0,N_1,\cdots,N_L\}$. The NN has $L$ \emph{junctions} in between the layers, with $N_{i-1}$ and $N_i$ respectively being the number of neurons in the earlier (left) and later (right) layers of junction $i$. Every left neuron has a fixed number of edges (or weights) going from it to the right, and every right neuron has a fixed number of edges coming into it from the left. These numbers are defined as out-degree ($\outdeg_i$) and in-degree ($\indeg_i$), respectively. For FC layers, $\outdeg_i = N_i$ and $\indeg_i = N_{i-1}$. In contrast, pre-defined sparsity leads to sparsely connected (SC) layers, where $\outdeg_i < N_i$ and $\indeg_i < N_{i-1}$, such that $N_{i-1}\times \outdeg_i = N_i\times \indeg_i = W_i$, which is the total number of weights in junction $i$. Having a fixed $\outdeg_i$ and $\indeg_i$ ensures that all neurons in a junction contribute equally and none of them get disconnected, since that would lead to a loss of information. The connection density in junction $i$ is given as $W_i/(N_{i-1}N_i)$ and the overall connection density of the network is defined as $\left( \sum_{i=1}^{L}{W_i} \right) /\ \left( \sum_{i=1}^{L}{N_{i-1}N_i} \right )$. Previous works \cite{Dey2017_ICANN,Dey2018_ITA} have shown that overall density levels of $<10\%$ incur negligible performance degradation -- which motivates us to implement such low density networks on hardware in the present work.

\subsection{Hardware Architecture}\label{hw_desc}
This subsection describes the mathematical algorithm and the subsequent hardware architecture for a NN using pre-defined sparsity. The input layer, i.e. the leftmost, is fed \emph{activations} ($a_0$) from the input data. For an image classification problem, these are image pixel values. Then the \emph{feedforward (FF)} operation proceeds as described in eq. \eqref{eq-ff}:
\begin{IEEEeqnarray}{c}\label{eq-ff}
a_i^{(j)} = \sigma \left( \sum _{f=1}^{\indeg_i} { w_{i}^{(j,k_f)}a_{i-1}^{(k_f)} + b_i^{(j)} } \right) \IEEEyesnumber \IEEEyessubnumber \label{eq-ff_a} \\
{\dot {a}}_i^{(j)} = {\sigma}^{'} \left( \sum _{f=1}^{\indeg_i} { w_{i}^{(j,k_f)}a_{i-1}^{(k_f)} + b_i^{(j)} } \right) \IEEEyessubnumber \label{eq-ff_b}
\end{IEEEeqnarray}
Both eqs. \eqref{eq-ff_a} and \eqref{eq-ff_b} are $\forall j \in \{1,\cdots,N_i\}, \forall i \in \{1,\cdots,L\}$. Here, $a$ is activation, ${\dot {a}}$ is its derivative (a-dot), $b$ is bias, $w$ is weight, and $\sigma$ and ${\sigma}^{'}$ are respectively the activation function and its derivative (with respect to its input), which are described further in Section \ref{fpga}. For $a$, ${\dot {a}}$ and $b$, subscript denotes layer number and superscript denotes a particular neuron in a layer. For the weights, ${w}_{i}^{(j,k_f)}$ denotes the weight in junction $i$ which connects neuron $k_f$ in layer $i-1$ to neuron $j$ in layer $i$. The summation for a particular right neuron $j$ is carried out over all $\indeg_i$ weights and left neuron activations which connect to it, i.e. $k_f \in \{1,\cdots,N_{i-1}\}$. These left indexes are arbitrary because the weights in a junction are \emph{interleaved}, or permuted. This is done to ensure good \emph{scatter}, which has been shown to enhance performance \cite{Dey2018_ITA}.

The output layer activations $a_L$ are compared with the ground truth labels $y$ which are typically one-hot encoded, i.e. $y^{(j)}$, $\forall j \in \{1,\cdots,N_L\}$, is 1 if the class represented by output neuron $j$ is the true class of the input sample, otherwise 0. We use the cross-entropy cost function for optimization, the derivative of which with respect to the activations is $a_L-y$. We also experimented with quadratic cost, but its performance was inferior compared to cross-entropy. The \emph{backpropagation (BP)} operation proceeds as described in eq. \eqref{eq-bp}:
\begin{IEEEeqnarray}{c}\label{eq-bp}
\delta_L^{(j)} = a_L^{(j)}-y^{(j)} \IEEEyesnumber \IEEEyessubnumber \label{eq-bp_a} \\
\delta_i^{(j)} = {\dot {a}}_{i}^{(j)} \left( \sum _{f=1}^{\outdeg_i} { w_{i+1}^{(k_f,j)}{\delta}_{i+1}^{(k_f)} } \right) \IEEEyessubnumber \label{eq-bp_b}
\end{IEEEeqnarray}
where $\delta$ denotes delta value. Eq. \eqref{eq-bp_a} is $\forall j \in \{1,\cdots,N_L\}$, and eq. \eqref{eq-bp_b} is $\forall j \in \{1,\cdots,N_i\}, \forall i \in \{1,\cdots,L-1\}$. The summation for a particular left neuron $j$ is carried out over all $\outdeg_i$ weights and right neuron deltas which connect to it, i.e. $k_f \in \{1,\cdots,{N}_{i+1}\}$. The right indexes are arbitrary due to interleaving.

Based on the $\delta$ values, the trainable weights and biases have their values updated and the network learns. We used the gradient descent algorithm, so the \emph{update (UP)} operation proceeds as described in eq. \eqref{eq-up}:
\begin{IEEEeqnarray}{c}\label{eq-up}
b_i^{(j)} \leftarrow b_i^{(j)} - \eta {\delta}_i^{(j)} \IEEEyesnumber \IEEEyessubnumber \label{eq-up_a} \\
w_{i}^{(j,k)} \leftarrow w_{i}^{(j,k)} - \eta a_{i-1}^{(k)}{\delta}_i^{(j)} \IEEEyessubnumber \label{eq-up_b}
\end{IEEEeqnarray}
where $\eta$ is the learning rate hyperparameter. Both eqs. \eqref{eq-up_a} and \eqref{eq-up_b} are $\forall i \in \{1,\cdots,L\}$. While eq. \eqref{eq-up_a} is $\forall j \in \{1,\cdots,N_i\}$, eq. \eqref{eq-up_b} is only for those $j \in \{1,\cdots,N_i\}$ and $k \in \{1,\cdots,N_{i-1}\}$ which are connected by a weight $w_{i}^{(j,k)}$.

\begin{figure}[!t]
\centering
\includegraphics[width = 0.7\linewidth]{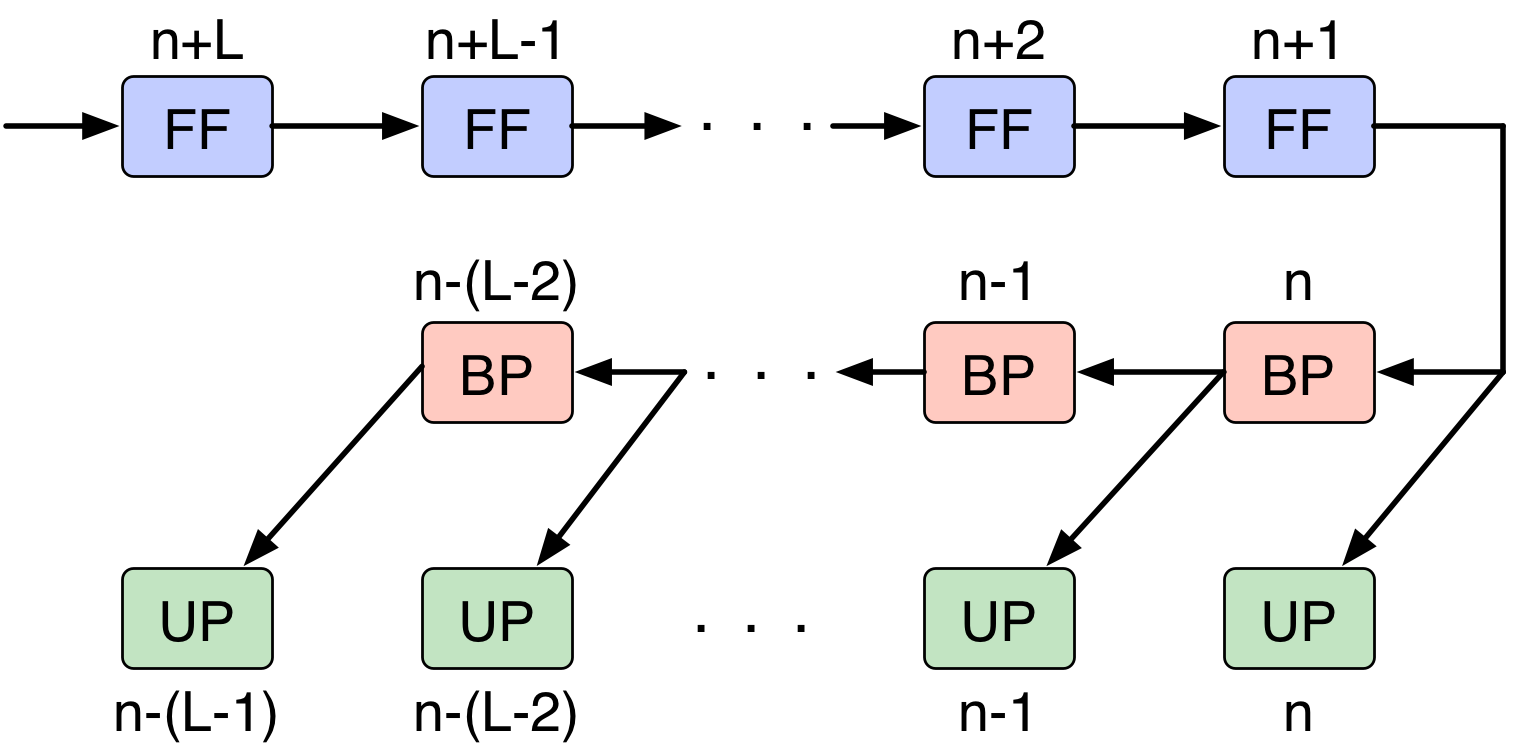}
\caption{Junction pipelining and operational parallelization in the architecture.}
\label{fig-jnpipelining}
\end{figure}

The architecture uses a) \emph{operational parallelization} to make FF, BP and UP occur simultaneously in each junction, and b) \emph{junction pipelining} wherein all the junctions execute all 3 operations simultaneously on different inputs. Thus, there is a factor of $3L$ speedup as compared to doing 1 operation at a time, albeit at the cost of increased hardware resources. Fig. \ref{fig-jnpipelining} shows the architecture in action. As an example, consider $L=2$, i.e. the network has an input layer, a single hidden layer, and an output layer. When the second junction is doing FF and computing cost on input $n+1$, it is also doing BP on the previous input $n$ which just finished FF, as well as updating (UP) its parameters from the finished cost computation results of input $n$. Simultaneously, the first junction is doing FF on the latest input $n+L = n+2$, and UP using the finished BP results of input $n-(L-1) = n-1$. BP does not occur in the first junction because there are no $\delta_0$ values to be computed.

The architecture uses \emph{edge processing} by making every junction have a \emph{degree of parallelism} $z_i$, which is the number of weights processed in parallel in 1 clock cycle (or simply cycle) by all 3 operations. So the total number of cycles to process a junction is $W_i/z_i$ plus some additional cycles for memory accesses. This comprises a \emph{block cycle}, the reciprocal of which is ideal throughput (inputs processed per second).

\begin{figure}[!t]
\centering
\includegraphics[width=0.7\linewidth]{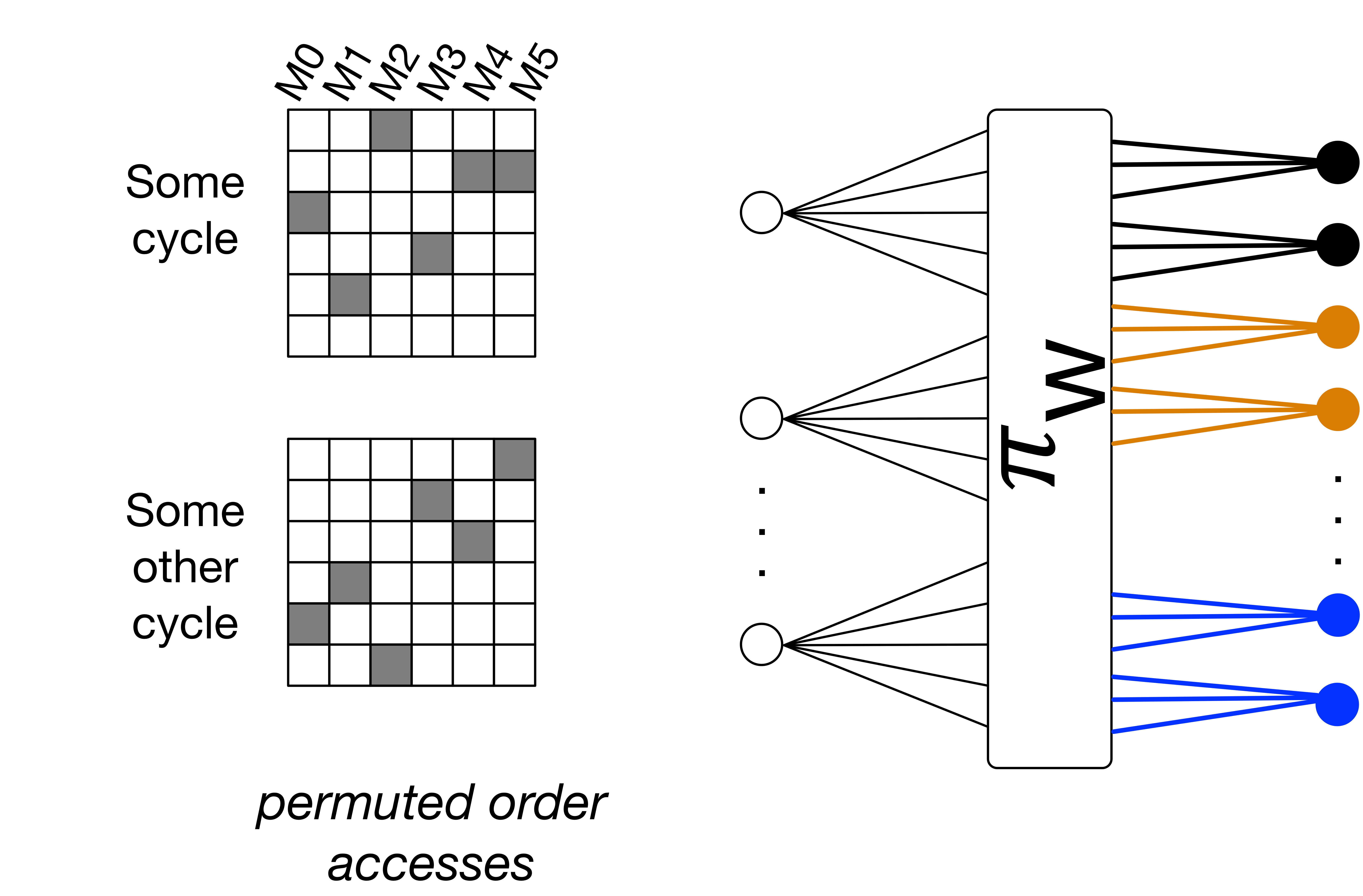}
\caption{Example of clash-freedom in some junction with $z=6$. In each cycle, $z$ weights are read corresponding to 2 right neurons (shown in same color). When traced back through the interleaver $\pi_W$, this requires accessing $z$ left activations in permuted order. There are $z$ activation memories $M0-M5$, only 1 element from each is read in a cycle in order to preserve clash-freedom. This is shown by the checkerboards, where only 1 cell in each column is shaded. Picture taken from \cite{Dey2017_Asilomar} with permission.}
\label{fig-interleaver_clashfreedom}
\end{figure}

\begin{figure}[!t]
\centering
\includegraphics[width = 0.9\linewidth]{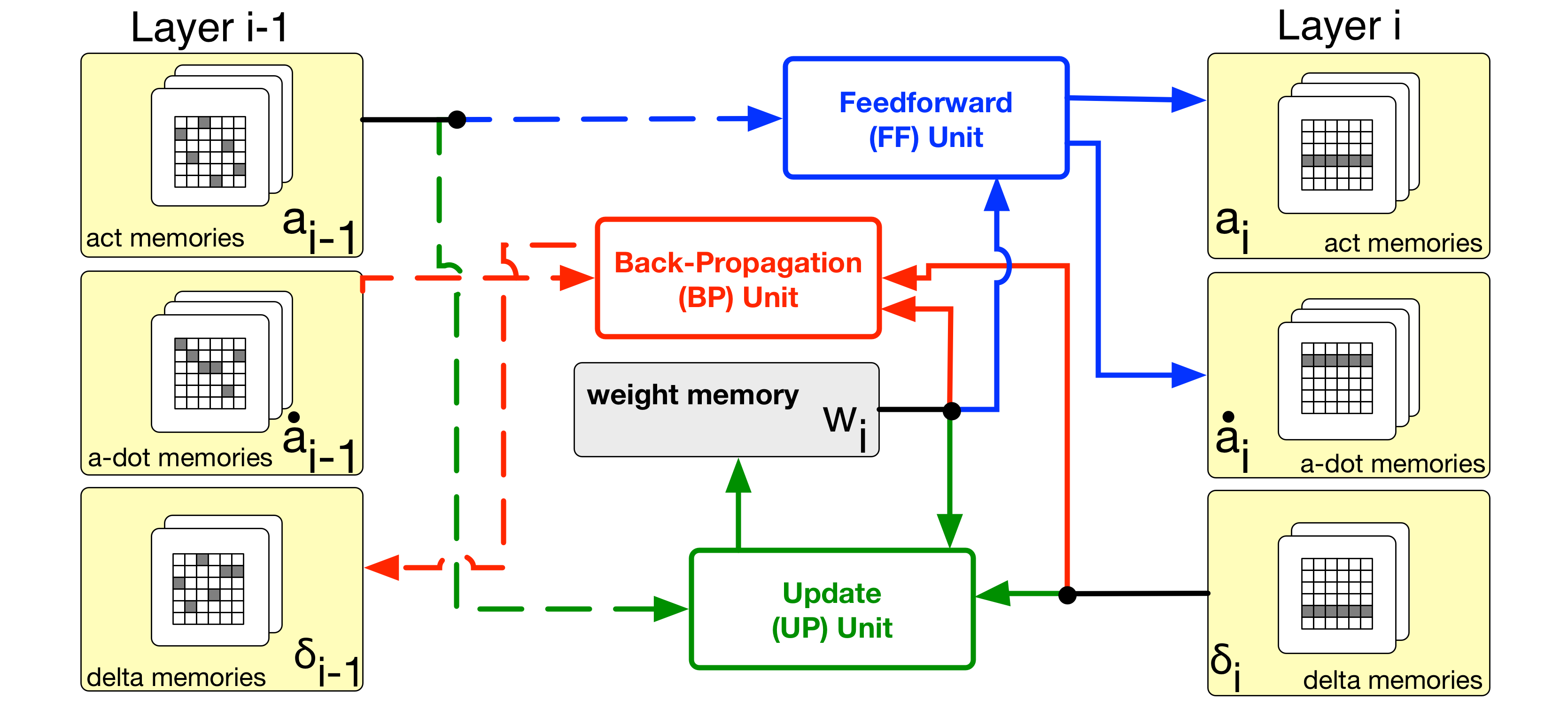}
\caption{Operational parallelization in junction $i$ ($i \ne 1$), showing natural and permuted order accesses as solid and dashed lines, respectively.}
\label{fig-pnp}
\end{figure}

All parameters and computed values in a junction are stored in banks of $z_i$ memories. The $z_i$ weights in the $k$th cells of all $z_i$ weight memories are read out in the $k$th cycle. 
Additionally, up to $z_i$ activations, a-dots, deltas and biases are accessed in a cycle. The order of accessing them can be natural (row-by-row like the weights), or permuted (due to interleaving). All accesses need to be \emph{clash-free}, i.e. the different values to be accessed in a cycle must all be stored in different memories so as to avoid memory stalls, as shown in Fig. \ref{fig-interleaver_clashfreedom}. 
Optimum clash-free interleaver designs are discussed in \cite{Dey2017_Asilomar}. Fig. \ref{fig-pnp} shows simultaneous FF, BP and UP, along with memory accesses, in more detail inside a single junction.

This architecture is ideal for implementation on reconfigurable hardware due to a) its parallel and pipelined nature, b) its low memory footprint due to sparsity, and particularly c) the degree of parallelism $z_i$ parameters, which can be tuned to efficiently utilize available hardware resources, as described in Sections \ref{impl} and \ref{effects_z}. 

\section{FPGA Implementation}\label{fpga}

\subsection{Device and Dataset}\label{board}
We implemented the architecture described in Section \ref{hw_desc} on an Artix-7 FPGA. 
This is a relatively small FPGA and therefore allowed us to explore efficient design styles and optimize our RTL to make it more robust and scalable. We experimented on the MNIST dataset where each input is an image consisting of 784 pixels in 8-bit grayscale each. Each ground truth output is one-hot encoded between 0-9. Our implementation uses powers of 2 for network parameters to simplify the hardware realization. Accordingly we padded each input with 0s to make it have 1024 pixels. The outputs were padded with 0s to get 32-bit one-hot encoding. Prior to hardware implementation, software experiments showed that having extra always-0 I/O did not detract from network performance.

\subsection{Network Configuration and Training Setup}\label{config}
The network has 1 hidden layer of 64 neurons, i.e. 2 junctions overall. Other parameters were chosen on the basis of hardware constraints and experimental results, which are described in Sections \ref{bitwidth} and \ref{impl}. The final network configuration is given in Table \ref{table-config}.
\begin{table}[!t]
\begin{minipage}{\columnwidth}
\renewcommand{\arraystretch}{1.2}
\caption{Implemented Network Configuration}
\label{table-config}
\centering
\begin{tabular}{|c|c|c|}
\hline
Junction Number ($i$) & 1 & 2\\
\hline
Left Neurons ($N_{i-1}$) & 1024 & 64\\
\hline
Right Neurons ($N_i$) & 64 & 32\\
\hline
Fan-out ($\outdeg_i$) & 4 & 16\\
\hline
Weights ($W_i=N_{i-1}\times \outdeg_i$) & 4096 & 1024\\
\hline
Fan-in ($\indeg_i=W_i/N_i$) & 64 & 32\\
\hline
$z_i$ & 128 & 32\\
\hline
Block cycle ($W_i/z_i$) \footnote{In terms of number of clock cycles. Not considering the additional clock cycles needed for memory accesses.} & 32 & 32\\
\hline
Density ($W_i/(N_{i-1}N_i)$) & 6.25\% & 50\%\\
\hline
Overall Density & \multicolumn{2}{c|}{7.576\%}\\
\hline
\end{tabular}
\end{minipage}
\end{table}

We selected $12544$ MNIST inputs to comprise 1 epoch of training. 
Learning rate ($\eta$) is initially $2^{-3}$, halved after the first 2 epochs, then after every 4 epochs until its value became $2^{-7}$. Dynamic adjustment of $\eta$ leads to better convergence, while keeping it to a power of 2 leads to the $\eta$ multiplications in eq. \eqref{eq-up} getting reduced to bit shifts. Pre-defined sparsity leads to a total number of trainable parameters $= \left(w_1=4096\right)+\left(w_2=1024\right)+\left(b_1=N_1=64\right)+\left(b_2=N_2=32\right) = 5216$, which is much less than $12544$, so we theorized that overfitting was not an issue. We verified this using software simulations, and hence did not apply weight regularization.

\subsection{Bit Width Considerations}\label{bitwidth}

\subsubsection{Parameter Initialization}
We initialized weights using the Glorot Normal technique, i.e. their values are taken from Gaussian distributions with mean $=0$ and variance $=2/\left(\outdeg_i+\indeg_i\right)$. This translates to a three standard deviation range of $\pm0.51$ for junction 1 and $\pm0.61$ for junction 2 in our network configuration described in Table \ref{table-config}.

\begin{figure}[!t]
\centering
\includegraphics[width = 0.7\linewidth]{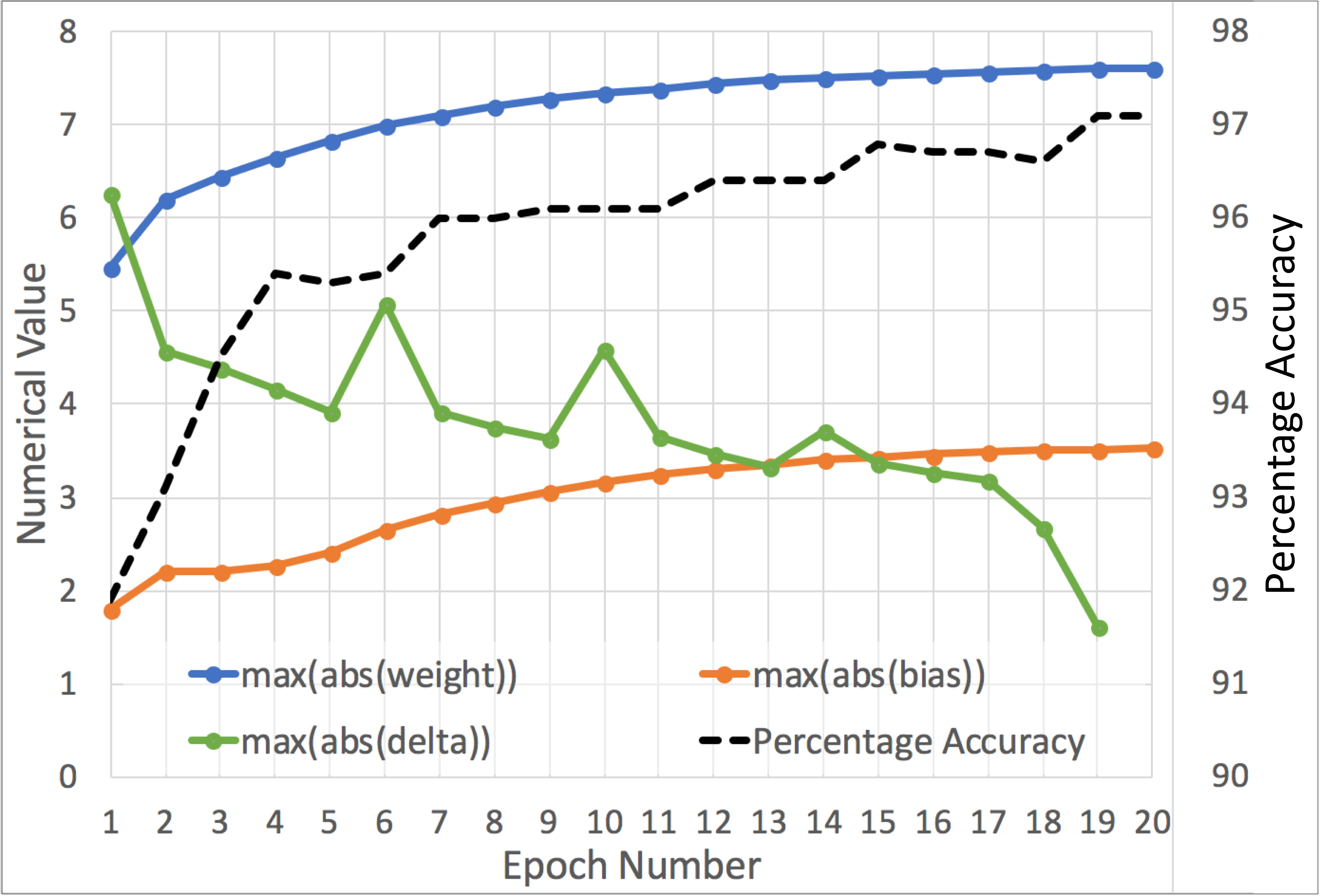}
\caption{Maximum absolute values (left y-axis) for $w$, $b$ and $\delta$, and percentage classification accuracy (right y-axis), as the network is trained.}
\label{fig-valueranges}
\end{figure}

The biases in our architecture are stored along with the weights as an augmentation to the weight memory banks. So we initialized biases in the same manner as weights. Software simulations showed that this led to no degradation in performance from the conventional method of initializing biases with 0s. This makes sense since the maximum absolute value from initialization is much closer to 0 than their final values when the network converges, as shown in Fig. \ref{fig-valueranges}.

To simplify the RTL, we used the same set of $W_i/z_i$ unique values to initialize all weights and biases in junction $i$. Again, software simulations showed that this led to no degradation in performance as compared to initializing all of them randomly. This is not surprising since an appropriately high value of initial learning rate will drive each weight and bias towards its own optimum value, regardless of similar values at the start.

\subsubsection{Fixed Point Configuration}
We recreated the aforementioned initial conditions in software and trained our configuration to study the range of values for network variables until convergence. The results for $w$, $b$ and $\delta$ are in Fig. \ref{fig-valueranges}. The $a$ values are generated using the \emph{sigmoid} activation function, which has range $= [0,1]$.

To keep the hardware optimal, we decided on the same \emph{fixed point} bit configuration for all computed values and trainable parameters --- $a$, $\dot{a}$, $\delta$, $w$ and $b$. Our configuration is characterized by the bit triplet $\left(b_w,b_n,b_f\right)$, which are respectively the total number of bits, integer bits, and fractional bits, with the constraint $b_w = b_n+b_f+1$, where the 1 is for the sign bit. This gives a numerical range of $[-{2}^{b_n},2^{b_n}-2^{-b_f}]$ and precision of $2^{-b_f}$. Fig. \ref{fig-valueranges} shows that the maximum absolute values of various network parameters during training stays within 8. Accordingly we set $b_n=3$. We then experimented with different values for the bit triplet and obtained the results shown in Table \ref{table-bitwidth}. Accuracy is measured on the last 1000 training samples. Noting the diminishing returns and impractical utilization of hardware resources for high bit widths, we chose the bit triplet $\left(12,3,8\right)$ as being the optimal case.

\begin{table}[!t]
\renewcommand{\arraystretch}{1.0}
\caption{Effect of Bit Width on Performance}
\label{table-bitwidth}
\centering
\begin{tabular}{|c|c|c|c|c|c|}
\hline
$b_w$ & $b_n$ & $b_f$ & FPGA LUT & Accuracy after & Accuracy after\\
 & & & Utilization \% & 1 epoch & 15 epochs\\
\hline
8 & 2 & 5 & 37.89 & 78 & 81\\
\hline
10 & 2 & 7 & 72.82 & 90.1 & 94.9\\
\hline
10 & 3 & 6 & 63.79 & 88 & 93.8\\
\hline
12 & 3 & 8 & 83.38 & 90.3 & 96.5\\
\hline
16 & 4 & 11 & \textcolor{red}{112} & 91.9 & 96.5\\
\hline
\end{tabular}
\end{table}

\subsubsection{Dynamic Range Reduction due to Sparsity}
\begin{figure}[!t]
\centering
\includegraphics[width = 0.6\linewidth]{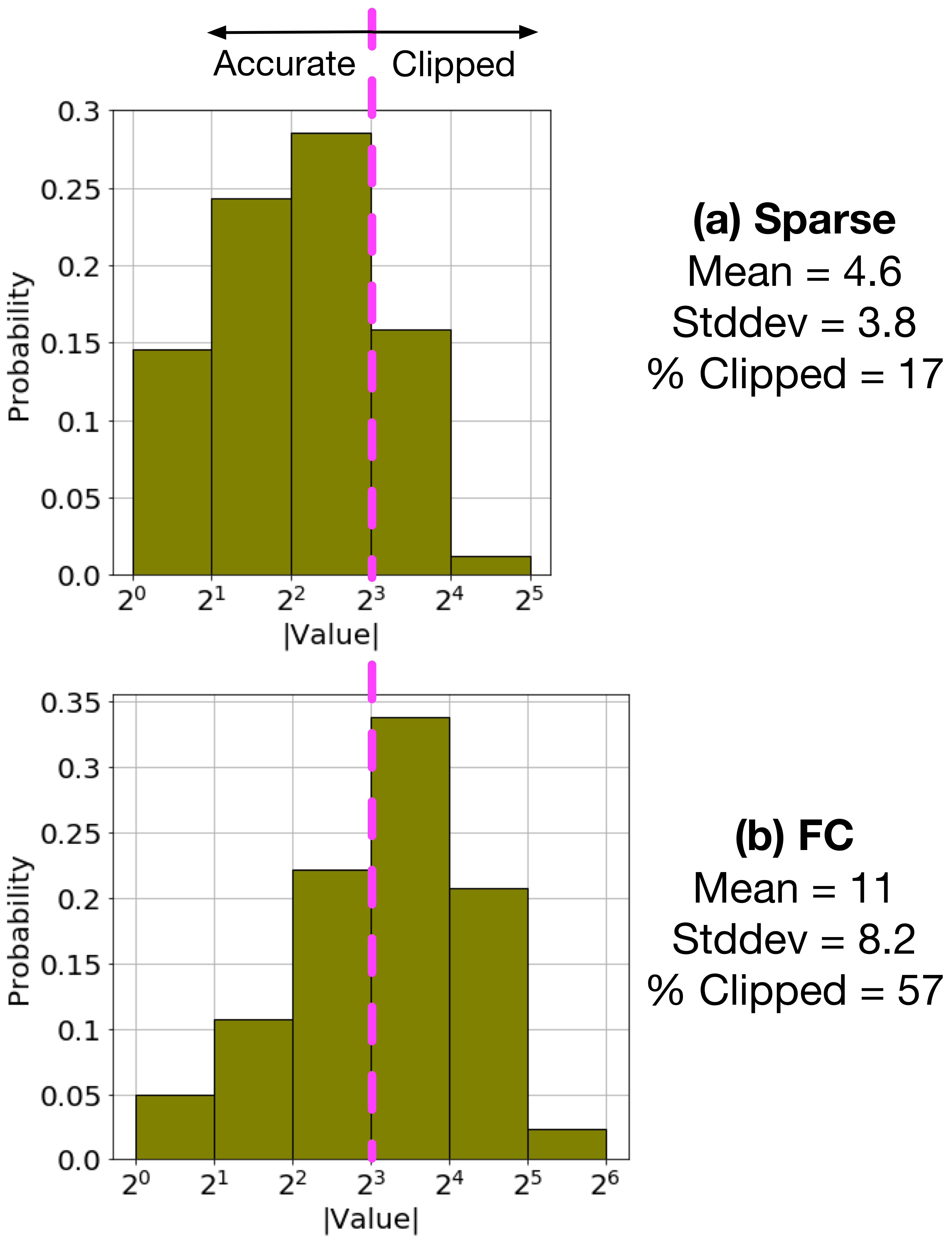}
\caption{Histograms of absolute value of eq. \eqref{eq-ff}'s $\sum{w_1a_0}+b_1$ with respect to dynamic range for (a) sparse vs. (b) FC cases, as obtained from ideal floating point simulations on software. Values right of the pink line are clipped.}
\label{fig-a1distribution}
\end{figure}

We found that sparsity leads to reduction in the dynamic range of network variables, since the summations in eqs. \eqref{eq-ff} and \eqref{eq-bp} are over smaller ranges. This motivated us to use a special form of adder and multiplier which preserves the bit triplet between inputs and outputs by clipping large absolute values of output to either the positive or negative maximum allowed by the range. For example, 10 would become 7.996 and $-10$ would become $-8$. 
Fig. \ref{fig-a1distribution} analyzes the worst clipping errors by comparing the absolute values of the argument of the sigmoid function in the hidden layer, i.e. $\sum{w_1a_0}+b_1$ from eq. \eqref{eq-ff}, for our sparse case vs. the corresponding FC case ($\outdeg_1=64$, $\outdeg_2=32$). Notice that the sparse case only has 17\% of its values clipped due to being outside the dynamic range afforded by $b_n=3$, while the FC case has 57\%. The sparse case also has a smaller variance. This implies that the hardware errors introduced due to finite bit-width effects are less pronounced for our pre-defined sparse configuration as compared to FC.

\subsubsection{Experiments with ReLU}
\begin{figure}[!t]
\centering
\includegraphics[width = 0.65\linewidth]{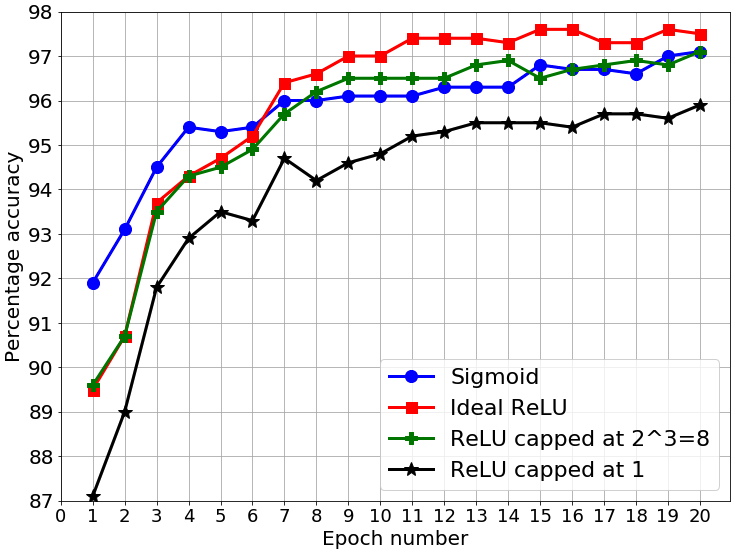}
\caption{Comparison of activation functions for $a_1$.}
\label{fig-act}
\end{figure}

As demonstrated in literature \cite{Krizhevsky2012,Szegedy2015,He2016}, the native (ideal) ReLU activation function is more widely used than sigmoid due to the former's better performance, no vanishing gradient problem, and tendency towards generating sparse outputs. However, ideal ReLU is not practical for hardware due to its unbounded range. We experimented with a modified form of the \emph{ReLU} activation function where the outputs were clipped to a) 8, which is the maximum supported by $b_n=3$, and b) 1, to preserve bit width consistency in the multipliers and adders and ensure compatibility with sigmoid activations. Fig. \ref{fig-act} shows software simulations comparing sigmoid with these cases. Note that ReLU clipped at 8 converges similar to sigmoid, but sigmoid has better initial performance. Moreover, there is no need to promote extra sparsity by using ReLU because our configuration is already sparse, and sigmoid does not suffer from vanishing gradient problems because of the small range of our inputs. We therefore concluded that sigmoid activation for all layers is the best choice. 

\subsection{Implementation Details}\label{impl}

\subsubsection{Sigmoid Activation}
The sigmoid function uses exponentials, which are computationally infeasible to obtain in hardware. So we pre-computed the values of $\sigma(\cdot)$ and ${\sigma}^{'}(\cdot)$ and stored them in look-up tables (LUTs). Interpolation was not used, instead we computed sigmoid for all 4096 possible 12-bit arguments up to the full 8 fractional bits of accuracy. On the other hand, its derivative values were computed to $6$ fractional bits of accuracy since they have a range of $[0,2^{-2}]$. Note that clipped ReLU activation uses only comparators and needs no LUTs. However, the number of sigmoid LUTs required is $\sum_{i=1}^{L}{z_i/\indeg_i}=3$, which incurs negligible hardware cost. This reinforces our decision to use sigmoid instead of ReLU.

\subsubsection{Interleaver}
We used clash-free interleavers of the \emph{SV+SS} variation, as described in \cite{Dey2017_Asilomar}. Starting vectors for all sweeps were pre-calculated and hard-coded into FPGA logic.

\subsubsection{Arithmetic Units}
We numbered the weights sequentially on the right side of every junction, which leads to permuted numbering on the left side due to interleaving. We chose $z_i\geq \indeg_i, \forall i \in \{1,\cdots L\}$. This means that the $z_i$ weights accessed in a cycle correspond to an integral ($z_i/\indeg_i$) number of right neurons, so the FF summations in eq. \eqref{eq-ff} can occur in a single cycle. This eliminates the need for storing FF partial sums. The total number of multipliers required for FF is $\sum _{i=1}^{L}{z_i}$. The summations also use a tree adder of depth $={\text{log}}_{2}\left(\indeg_i\right)$ for every neuron processed in a cycle.

BP does not occur in the first junction since the input layer has no $\delta$ values. The BP summation in eq. \eqref{eq-bp_b} will need several cycles to complete for a single left neuron since weight numbering is permuted. This necessitates storing $\sum _{i=2}^{L}{z_i}$ partial sums, however, tree adders are no longer required. Eq. \eqref{eq-bp_b} for BP has 2 multiplications, so the total number of multipliers required is $2\sum _{i=2}^{L}{z_i}$.

The UP operation in each junction $i$ requires $z_i$ adders for the weights and $z_i/\indeg_i$ adders for the biases, since that many right neurons are processed every cycle. Only the weight update requires multipliers, so their total number is $\sum _{i=1}^{L}{z_i}$.

Our FPGA device has 240 DSP blocks. Accordingly, we implemented the 224 FF and BP multipliers using 1 DSP for each, while the other 160 UP multipliers and all adders were implemented using logic.

\subsubsection{Memories and Data}
All memories were implemented using block RAM (BRAM). The memories for $a$ and $\dot{a}$ never need to be read from and written into in the same cycle, so they are single-port. $\delta$ memories are true dual-port, i.e. both ports support reads and writes. This is required due to the read-modify-write nature of the $\delta$ memories since they accumulate partial sums. The `weight+bias' memories are simple dual-port, with 1 port used exclusively for reading the $k$th cell in cycle $k$, and the other for simultaneously writing the $(k-1)$th cell. These were initialized using Glorot normal values while all other memories were initialized with 0s.

The ground truth one-hot encoding for all $12544$ inputs were stored in a single-port BRAM, and initialized with word size $=10$ to represent the 10 MNIST outputs. After reading, the word was padded with 0s to make it 32-bit long. On the other hand, the input data was too big to store on-chip. Since the native MNIST images are $28\times28=784$ pixels, the total input data size is $12544\times784\times8 = 78.68$ Mb, while the total device BRAM capacity is only $4.86$ Mb. So the input data was fed from PC using UART interface.

\subsubsection{Network Configuration}
\begin{figure}[!t]
\centering
\includegraphics[width = 0.7\linewidth]{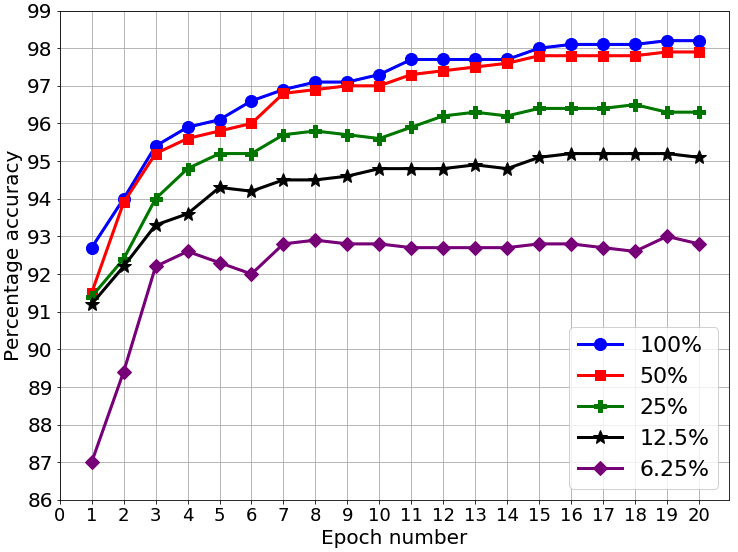}
\caption{Performance for different junction 2 densities, keeping junction 1 density fixed at 6.25\%.}
\label{fig-jn2density}
\end{figure}

Here we explain the choice of network configuration in Table \ref{table-config}. We initially picked $N_2=16$, which is the minimum power of 2 above 10. Since later junctions need to be denser than earlier ones to optimize performance \cite{Dey2018_ITA}, we experimented with junction 2 density and show its effects on network performance in Fig. \ref{fig-jn2density}. We concluded that 50\% density is optimum for junction 2. Note that individual $z_i$ values should be adjusted to have the same block cycle length for all junctions. This ensures an always full pipeline and no stalls, which can achieve the ideal throughput of 1 input per block cycle. This, along with the constraint $z_i\geq \indeg_i, \forall i \in \{1,\cdots L\}$, led to $z_1=256$, which was beyond the capacity of our FPGA. So we increased $N_2$ to 32 and set $z_2$ to the minimum value of 32, leading to $z_1=128$. We experimented with $\outdeg_1=8$, but the resulting accuracy was within 1 percentage point of our final choice of $\outdeg_1=4$.

\subsubsection{Timing and Results}
A block cycle in our design is $\left(W_i/z_i+2\right)$ clock cycles since each set of $z_i$ weights in a junction need a total of 3 clock cycles for each operation. The first and third are used to compute memory addresses, while the second performs arithmetic computations and determines our clock frequency, which is 15MHz.

We stored the results of several training inputs and fed them out to 10 LEDs on the board, each representing an output from 0-9. The FPGA implementation performed according to RTL simulations and within $1.5$ percentage points of the ideal floating point software simulations, giving 96.5\% accuracy in 14 epochs of training.

\subsection{Effects of $z$}\label{effects_z}

\begin{figure}[!t]
\centering
\includegraphics[width = 0.95\linewidth]{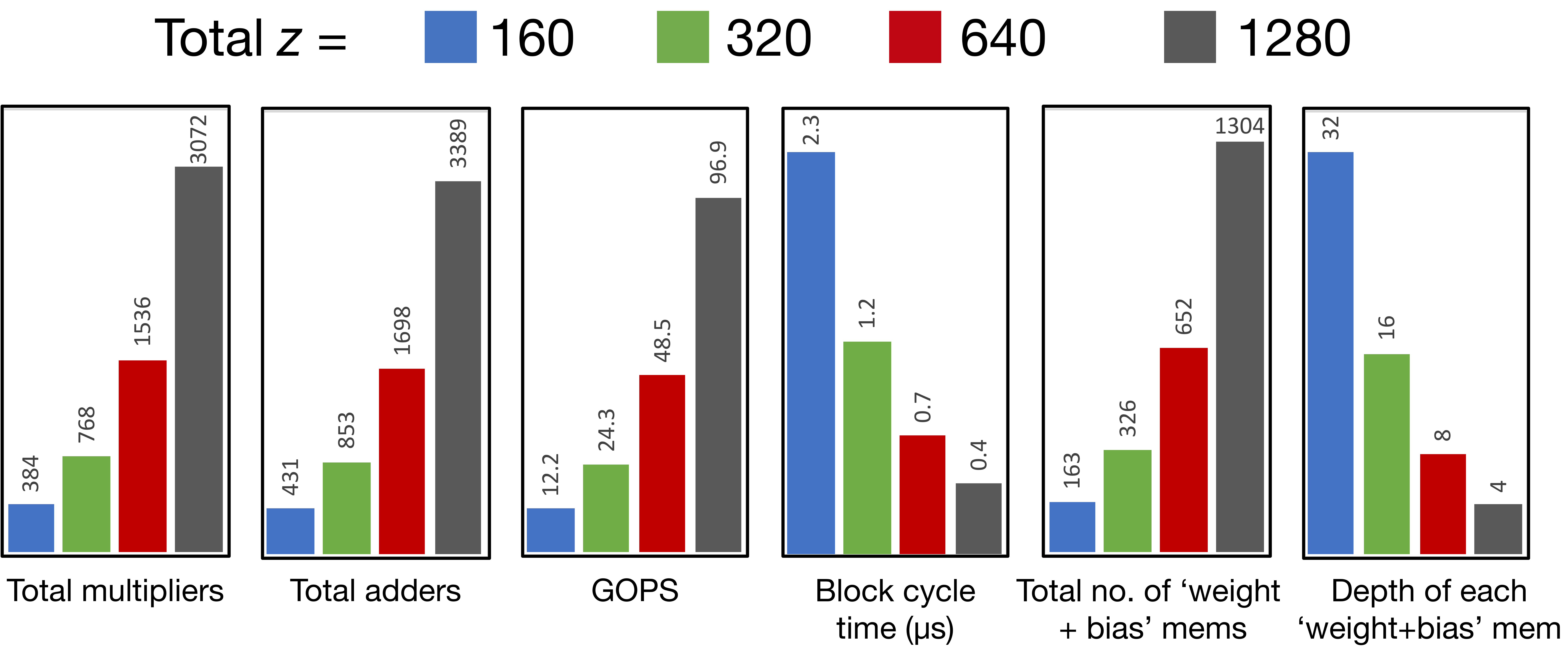}
\caption{Dependency of various design and performance parameters on the total $z$, keeping the network architecture and sparsity level fixed.}
\label{fig-effects_z}
\end{figure}

A key highlight of our architecture is the total degree of parallelism $\sum_{i=1}^{L}{z_i}$, which can be reconfigured to trade off training time and hardware resources while keeping the network architecture the same. This is shown in Fig. \ref{fig-effects_z}. The present work uses total $z=160$, which leads to a block cycle time of $2.27\mu s$, but economically uses arithmetic resources and has a small number of deep memories, making it ideal for a fully BRAM implementation. Given more powerful FPGAs, the same architecture can be reconfigured to achieve higher GOPS count and process inputs in $0.4\mu s$, albeit at the cost of more FPGA resources and a greater number of shallower memories. 
Moreover, this reconfigurability also allows a complete change in network structure and hyperparameters to process a new dataset on the same device if desired.

\section{Conclusion}\label{conc}
This paper demonstrates an FPGA implementation of both training and inference of a neural network pre-defined to be sparse. The architecture is optimized for FPGA implementation and uses parallel and pipelined processing to increase throughput. The major highlights are the degrees of parallelism $z_i$, which can be quickly reconfigured to re-allocate FPGA resources, thereby adapting any problem to any device. While the present work uses a modest FPGA board as proof-of-concept, this reconfigurability is allowing us to explore various types of networks on bigger boards as future work. Our RTL is fully parametrized and the code available 
on request.

\bibliographystyle{IEEEtran}
\bibliography{IEEEabrv,aaa_main}

\end{document}